# A "Participatory Universe" of J. A. Wheeler as an Intentional Correlate of Embodied Subjects and an Example of Purposiveness in Physics[1]

Alexei V. Nesteruk[2]
Department of Mathematics, University of Portsmouth, Lion Gate Building, Lion Terrace
Portsmouth PO1 3HF, UK

**Abstract**
This paper investigates the role of human subjectivity and its delimiters in articulating the universe in physics and cosmology. As a case study, we reflect upon the complex of ideas of the so called Participatory Universe by later J. A. Wheeler. The objective of the paper is to explicate the role of the human agency as a centre of disclosure and manifestation of the universe as well the as teleology of scientific representation of the world implied by the intrinsic purposiveness of human actions.

> A metaphysical interpretation and understanding of the world is neither scientifically attainable nor scientifically excluded. It is another mode of cognitive approach to the world, a transition from the (as much as possible) neutral observation of the world to a personal relationship with the world. It is a product of the freedom of humankind, and therefore interpretation and understanding define its entire stance towards the world , its mode of use of the world.
>
> Scientific observation does not simply affirm the reality of the cosmos; it constitutes it as an existential fact...,then every reality is recapitulated in the relationship of humanity with an active reason (*logos*) as an *invitation-to-relationship*, which is directed towards humanity alone.
>
> Christos Yannaras, *Postmodern Metaphysics*, pp. 114, 118, 137.

[2] E-mail: alexei.nesteruk@port.ac.uk

**Introduction**

In a recent paper (Nesteruk, 2012[2]) the issue of delimiters in cosmological research which originate in the structure of the human knower was addressing in particular how the purposiveness of human actions cascades towards the purposiveness of cosmological research (Nesteruk, 2012[3]). The latter paper dealt with a "formal" purposiveness in cosmology related to the explicability of the universe. This explicability is linked to the human intentional search for the sense of its own existence in the universe, so that the purpose of explanation in cosmology is related to the explication of the human condition. It was argued, in particular, that the theoretical representations of the "universe as a whole" and "the Big Bang" (as the encapsulated origin of the universe) act as the *telos* of cosmological explanation and, hence, as well, as the *telos* of anthropological explanation related to the origin of individual persons at birth (Nesteruk 2012[1]). In this paper we would like to discuss, as a case study, an interesting example of how scientific development in the 20th century led a famous physicist John Archibald Wheeler to extend the naturalistic methodology together with classical ideal of rationality (where subject and object are entirely separated and the world is supposed to pre-exist independently of human insight and its activity) towards that which can be described as a phenomenological stance portraying man as the centre of disclosure and manifestation of the world. It is of interest that this extension has some *teleological* connotations, bringing teleology into the heart of scientific explicability of the universe.

Wheeler, after a long intellectual evolution working in physics, attempted to approach physical reality not as something "out there", which is passively described by observers, but to see it as a genesis through conscious dialogue between observers-participants and physical reality, so that the universe emerges as a special articulation of the relationship between human intelligence and physical reality (Wheeler 1994[1], p. 128). This approach, challenging the natural scientific attitude, was not appreciated by physicists who found Wheeler's ideas "unpalatable in view of its rather mystical overtones" (Carr 1998, p. 158), and hence has not received any further attention and development. A sceptical reaction of physicists to Wheeler's ideas can be understood because his ideas represent a metaphysical extension of physics which physicists do not consider as a part of their vocation and duty. However, seen historically and in particular in conjunction with philosophical developments in the 20th century, this was not an entirely arbitrary attempt for it manifested a certain inevitability of sliding towards a transcendental or phenomenological appropriation of physics if the latter were to be to tackle the issue of its own foundation and its very facticity. A simple question which must be posed by any physicist who is interested to know the truth would this: why is physics possible at all? And here the question is not only about the intelligibility of the world, but rather of the very basic existential premises of physics related to humanity as its agent. Physics, as a science and social activity, is a product and a certain accomplishment of human beings who are themselves part of the physical world. In this sense the facticity of physics is related to a particular position of human beings in the world, such that this world allows them to produce its own explication and description. On purely philosophical grounds, this explicability and description has an absolutely contingent character related, speaking in Heideggerian terms, to that fragment of the unconcealed being which is associated to a specific living presence, that is human persons. Still, for physicists, prone to reductionism, there remains a question as to whether physics itself can explicate its own existence, or, in a slightly different parlance, can some simple initial rules of interaction with the world (which, in fact, presuppose the world's explicability from the beginning) lead with necessity to that picture of the world which we have here and now. In this sense the case of Wheeler's thought (in spite of deviating from the established stream of physics) represents an example, in the history of scientific ideas, of how a naturalistic epistemology in science in attempting to make a certain self-correction, through the search for its own facticity, leads to a transcendental problematic (remarkably with teleological overtones), that is to the view that the complete picture of physical reality must include the conditions of its explicability and constitution.

Wheeler develops his own transcendental argument basing himself in Einstein's theory of Relativity and Quantum Mechanics which, according to him, changed the vision of the human position in the universe by making human beings co-creators of physical reality in a very non trivial sense. He reformulates de facto the famous paradox of human subjectivity in the world[3] which states, on the one hand, that humanity communicates some palpability and sense to the physical world, and, on the other hand, the fact that human incarnate subjectivity is a finite accomplishment of this world:

"The brain is small. The universe is large. In what way, if any, is it, the observed, affected by man, the observer? Is the universe deprived of all meaningful existence in the absence of mind? Is it governed in its structure by the requirement that it gives birth to life and consciousness? Or is man merely an unimportant speck of dust in a remote corner of space? In brief, are life and mind irrelevant to the structure of the universe – or are they central to it?"(Wheeler 1975, p. 270).[4]

Let us comment on this passage from the point of view of the already existing insights which came before Wheeler in philosophy and theology.

"The brain is small. The universe is large." Indeed the size of the physical organ, which is responsible for mental articulation of the whole universe is incommensurable with the spatial size of the visible universe. Still, and this is an existential fact, it is from within this spatial scale that the articulation of microscopic realities of particles and fields, as well as huge astronomical formations is possible by this organ. There is something in this incommensurability, which is not physical, or, at least is not based on physical interactions. The very idea of a continuum of the universe as a single and united whole, although inaccessible to the empirical grasp, reflects a non-local and non-physical property of the world which is detected by consciousness through the power of intuition.[5]

[3] This paradox is a perennial problem of philosophy and was anticipated by ancient Greek philosophers and Christian thinkers. It was expressed differently by such philosophers as Kant (see, for example, Kant's conclusion to his *Critique of Practical Reason*.) Among phenomenological philosophers who dealt with this paradox one can mention E. Husserl, M. Scheler, M. Merleau-Ponty, E. Fromm and others. The general discussion of this paradox can be found in (Carr, 1999). The role of this paradox in discussions on science and theology can be found in (Nesteruk, 2008, pp. 173-75). Applied to the study of the universe the paradox of human subjectivity can be formulated as follows: on the one hand human beings in the facticity of their embodied condition form the centre of disclosure and manifestation of the universe as a whole, modelling it as overall-space and time which exceeds the limits of the attuned space related to humanity's comportment on the planet earth (the home place). On the other hand the depicted universe as a vast continuum of space and time positions humanity in an insignificant place in the whole totality making its existence not only contingent (in physical terms) but full of nonsense from the point of view of the actually infinite universe. Said bluntly the actual infinity of the universe is attempted to be articulated from an infinitely small part of its formation. One could express this differently: through its insight humanity is co-present in all points of what it observes in the universe, or imagines, while physically being restricted to an insignificant part of it.

[4] Certainly such a questioning on the place of humanity in the universe is not novel in history of thought and philosophy. It is enough to point to Pascal, who compared human being with the "thinking reed" whose position in the universe is ambivalent because of the physical insignificance and epistemological centrality: "Man is a reed, the feeblest in nature; but he is a thinking reed. The whole universe need not take up arms to crush him…But even if the universe should crush him, man would still be more noble than that which kills him, since he knows he is mortal, and knows that the universe is more powerful that he: but the universe itself knows nothing of it" (Pascal 1959, p. 78).

[5] See in this respect (Weyl 1994).

"In what way, if any, is it, the observed [the universe], affected by man, the observer? Is the universe deprived of all meaningful existence in the absence of mind?" Physics teaches us that, through our own spatial and temporal insignificance in the whole grandeur of the universe, we are just late newcomers into this world who only recently started to interfere with the physical environment on this planet. Our ability to affect the cosmos at large is only a matter of science fiction and some futuristic prophecies. [6] However the question of Wheeler has another, much more serious dimension related to epistemology: will the universe as we observe it, that is see it in its particular contingent appearance as intricately related to our physiological and psychological constitution, be unconcealed to us in a different way, related to that measure, which man will be, in relation to that which can be unconcealed. Responding to the second half of a question formulated in the beginning of this paragraph, the question of the universe in absence of the human mind is an ontic question: indeed one can build, so to speak onto-cosmology (in analogy with onto-theology criticized by Heidegger), in which the universe will be an impersonal being allegedly existing independently of the human grasp, to which one can attribute many intellectually imposed properties. The question then is what is the value of this universe for human beings. Rephrasing Heidegger, "can one dance and sing in front of such a universe"? (Heidegger 1969, p. 72). The answer has been given long before by the proponents of theological insights, that the universe without human beings is dumb and it is humanity which is the "voice" (hypostasis) of the universe (See, for example, (Torrance 2001, p. 4), (Clément1976, p. 91)). The question of Wheeler is exactly about this: can one predicate the universe as existent without regarding humanity in measure of interaction with which this universe comes to its unconcelment? The development of physics in the 20th century with its increasing understanding that its results depend on the contexts which are not strictly objective and detached from us, but are set up by us through experiments and measurements, led Wheeler to an intuition that the mechanistically constructed representation of the universe remains no more than an *idea*, a mental accomplishment. Correspondingly the "meaningful existence of the universe in the absence of the mind" is a contradiction in terms, for the very word meaning has strictly human connotations (if we avoid any references to theology, which can suggest that the meaning of the universe originates in the creative and willing activity of the divine agency, which or who sustains this universe through creation out of nothing). In view of this, Wheeler attempts to address the issue of meaning: where the meaning comes from and whether it can come from some underlying physics, initially free from human insight physics.

Wheeler's enquiry into the foundations of the historical contingent facticity of physics came to its explicit manifestation in his train of thought after a long period in his scientific activity when he was following a scientific programme of Einstein, who believed that it was possible to unify different physical forces reducing them to some geometrical effects. Wheeler spent a considerable effort to advance the so called "geometrodynamics" whose essence was also to explain the genesis of macroscopic space and time by appealing to some underlying structures which follow the rules of quantum physics (Wheeler 1968). Wheeler argued that space-time is a classical concept, an approximation, which is incompatible with the quantum principle (Wheeler 1973, p. 227). This meant to Wheeler that the basic ingredients of classical physical theories, such as space and time, cannot survive their extrapolation into the microscopic world where quantum principle rules, so that space and time are not basics and are subject to change and further explication. Generalising this observation Wheeler makes a much more radical conclusion about the

[6] One of such prophecies was promoted by F. Tipler in his book (Tipler 1995) where he develops an idea of such a large-scale affection of the universe which will guarantee the possibility of an indefinite information processing, which, according to Tipler mimics the persistence of life in the universe.

*mutability* of physics, which implies that all classical physical concepts, while they loose their sense in the limiting case of microscopic scales, and, in particular, in a context of gravitational collapse predicted by general relativity, are subject to *genesis* from the realm which has no obvious and visual physical characteristics at all. In other words, physics which is usually associated with some immutable laws, constants of nature and harmonies in the world ceases to function in extreme cases such as gravitational collapse corresponding to the origin of the universe, so that, according to Wheeler, "there is no law except the law that there is no law" or "ultimate mutability is the central feature of physics" (Wheeler 1973, p. 242). One must reflect upon this last assertion with a grain of a philosophical scepticism, for the ultimate mutability of physics cannot serve as its transcendental delimiter: physics is impossible if mutability reigns in the universe simply because experience cannot then be ordered. Mutability excludes any identity in time for, according to Wheeler, there is no time, so that the mutability implies an infinite degree of differentiation in being, which cannot be stabilised even hypothetically in any reflecting thought. Still this mutability has a limit: the world of existences with human observers must be produced from it and retrospectively constituted. Thus from the beginning the mutability affirmed has its limits originating in the counterfactual causality towards this mutability following from the actuality of the empirically present (and immutable to some extent) physical world. This causality is linked to the human agency which constitutes the world. And this human agency enters Wheeler's scheme through the "quantum principle" by which he means an epistemological, as well as ontological claim originating in the extreme version of the so called Copenhagen interpretation of quantum mechanics that quantum phenomenon is phenomenon as long as it is observed and articulated by some intelligible agency. This implies that what physics calls "nature" is not just something "out there", in itself, but that which is constituted through the interaction of intelligible beings (who are capable of asking questions and receiving the responses) with that "something" which is initially inarticulate and which is being questioned. Wheeler writes: "Nothing is more astonishing about quantum mechanics than its allowing one to consider seriously the view that the universe would be nothing without observership as surely as a motor would be dead without electricity" (Wheeler 1994, p. 39). Then he comments on observership, referring to the views of Bohr and Wigner who advocated that observation and measurement are complete when they enter consciousness of an observer and then can be communicated to another observer in a plain language; ".. an experiment is only an experiment when the outcome is expressed in the form of communicable knowledge, knowledge which can be shared" (Wheeler 1994[1], p. 26). But "observership" is not a simple term, it cannot be defined prior to the act of observation and establishing its meaning: "What is 'observership'? It is too early to answer. [..] The main point here is to have a word that is not defined and never will be defined until that day when one sees much more clearly [..] how the observations of all the participators, past, present and future, join together to define what we call 'reality' " (Wheeler 1994[1], p. 26). By employing a phenomenological language, reality, according to Wheeler, is defined as an intentional correlate of cumulative acts of observation and as a communal accomplishment along a particular, historically contingent, but, perhaps, teleologically driven ways. The meaning of reality can only be established if there is a field of intersubjectivity with some trans-empirical features, which transcend physical past, present and future. It is in the framework of this intersubjectivity through its continuous embodiment through observership-participation that the truth and *meaning* of things is established.

Being a physicist, Wheeler does not pretend to provide any sophisticated theory of how to derive the meaning of physical concepts such as space, time, particles etc. from a deep level of human subjectivity, that is the outline of their constitution. This is the work of phenomenologists who do not appeal to any vague physical analogies. Instead of this Wheeler traces the logic which is inherent in physical thought which makes the genesis of these concepts plausible through their consistency with each other. On the basis of this he conjectures that we enter now a "third era of physics", which should explain the genesis of such concepts: "we have to account for all the structure that makes physics what it is" (Wheeler 1983, p. 404).[7] Wheleer believes that the question "What makes meaning?" applied to physics, is an existential question, for it also addresses the issue of the existence of human beings and the universe. But unlike existential philosophers, who were sceptical about science's capacity to deal with this issue[8], he believes that physics itself can address the issue of the facticity of existence: "Tomorrow, will it not be existence itself that comes under the purview of physics?" (Wheeler 1983, p. 404).

The more radical, metaphysically oriented conclusion of Wheeler is that the *overall reality*, that is the *totality of the world*, is constituted through the interaction between the inarticulate "out there" with human intelligent agencies who create the network of questions and answers directed to and received from what they intend to call "reality". It is interesting that this trend of Wheeler's thought is similar to phenomenology which asserts nature, as articulated worldly reality, as having sense only in the context of the dialogue between human consciousness and that which is posited by consciousness. It is the dialogue with the unarticulated otherness of consciousness that ultimately reveals the meaning. J. Kockelmans, a philosopher and commentator of Husserl, writes with respect to this dialogue that the meaning of the world arises in the encounter between man and the world and "exists only in an interplay of question and answer. We find the question in the world but it is still implicit and vague. Through my reply, which itself is a question, the first question becomes sharper so that a more accurate answer becomes possible. Meaning arises in a dialectic relationship between man and the world, but it is not possible to say which of the two first begins the 'interplay' and which of the two first gives meaning to the other" (Kockelmans 1966, p. 53). This passage is strikingly similar to Wheeler's ideas that physical reality reveals itself as an evolving complex of meanings in the course of the interplay between questions and answers which the human subject addresses to and receives from that "out there" which is eventually constituted by human observers as the physical reality and nature. Wheeler writes: "Physics gives light and pressure-tools to query and to communicate. Physics also gives chemistry and biology and, through them, observer-participators. They, by way of the devices they employ, the questions they ask, and the registrations they communicate... develop all they know or ever can know about the world" (Wheeler 1988, p. 5). Elsewhere he develops a simple analogy with a game of twenty questions which aims to recognise a word preconceived and hidden by a person through a simple process of interrogation of this person subjected to a single rule that it must be consistent with all previous questions and responses. In fact, if this word was not preconceived in advance, it will inevitably be constituted through the logic of the game simply because there must be answers "yes" or "no" which through a certain logic and questions' content, constitute the sought word (Wheeler 1979). Similarly, he claims, in nature, by asking questions we initiate the process of nature's response, which, in the course of enquiry leads us to the constitution of that which we intuitively aimed. Phenomenology described this process as a mental accomplishment of what is understood by nature: "nature", which science aims at through idealization and mathematisation, is not something a-priori given to human observers and thinkers, but something which is constructed and evolved towards an indefinite *telos*. "Nature", thus constructed, becomes exteriorised as an ideal (Gurwitsch 1974, p. 46) which is subject to accomplishment in a historical movement of scientific research because mathematics as human science is far from being static and accomplished,[9] and its advance creates more space for physicists to invade the realm of the yet unknown (although, perhaps, intelligible and invisible). In this sense, in analogy with phenomenology which makes a distinction between nature as it appears in primary perceptual experience (the inarticulate out there in Wheeler's scheme) and nature-for-physicists (that is "nature" as constituted through questions and answers), which is a mental accomplishment, as an ideal limit of convergent sequences of "images of nature" which are constructed by physicists in the course of history, one should treat Wheeler's genesis of meaning as an ever-going mental completion of the concept of nature. In the words of another phenomenologist, A. Gurwitsch, "nature" appears to be a "hypostasis of mental creations"[10]: this terminology resembles an old philosophical and theological notion of the so called *en-hypostasization* (personification, or human articulation) of nature or the universe.[11] Seen in this perspective, Wheeler's treatment of physical reality through quantum and computational synthesis makes it clear that the notion of physical reality, nature, or the universe has sense only in the context of humanity, which is in a position to relate it to the commonalities of perceptual experience, that is to put it in the context of incarnate existence-in-situation, as well as recapitulate them through an intelligible representation. It is in this sense that one can suggest that all images of the mathematised nature manifest the presence of the immutable dualistic constitution of things in the world, namely a fundamental differentiation (Gr.: *diaphora*) between the empirical and intelligible.

[7] Wheeler's appeal is in a way similar to a phenomenological assertion that without taking into account the generating power of human subjectivity, the efficacy of the sciences, either human or natural, remains obscured; see e.g. (Gurwitsch 1966, pp. 399-400).

[8] Existentialists considered the fact of life as a primary data for any further philosophising, the fact which as such cannot be placed in any allegedly wider and more general framework of the world's necessities. The radical metaphysical mystery of existence is expressed in human inability to "look" at this existence from outside.

When we have pointed out that the construct of "physical nature" in Wheeler's scheme represents an ideal, which can only be accomplished in the whole of the historical process, we assumed this as a philosophical hypothesis about hidden teleology in the scientific advance. Correspondingly, the progress of articulation of nature through its computational synthesis has meaning only as one particular tendency of the human spirit under which scientific knowledge and technology advance. The universe then is to be constructed along some particular path of knowledge corresponding to the human condition. Indeed Wheeler's questions and answers create a particular way forward in bringing reality to unconcelament, the way which in its historical concreteness is contingent and non-generic. However, one must recognise together with a phenomenological critique of the mathematisation of nature[12], that the universe constructed along the lines of Wheeler's scheme represents characteristically the fragmentation of the primary existential link between humanity and the world, considered through a particular discursive function of the human intellect which is based on abstraction and idealisation. "Nature" in the thus understood scientific sense, being a particular human accomplishment, does not exhaust the totality of reality. On the one hand the constructed "nature" is exteriorised by human subjectivity and is intended as being devoid of its inward existence in the hypostasis of human beings, on the other hand the same "nature", as *being constructed*, still entails some traces of its hypostatic origin. In different words, "nature" appears in a mode of *intentional immanence* related to those aspects of the overall reality which are not hypostatic in themselves, but *en-hypostasised* by human obsrervers-participants. By taking de facto a phenomenological stance, Wheeler proclaims that the world is not a clock-like machine which has been pre-constructed and then discovered by human observers; it is a self-synthesized system, coming into existence through the constitution of reality via questions and answers processed by a collective of persons-observers who are capable of establishing the meaning and interpretation of their observation-participancy ultimately leading to an integral view of nature.

[9] See on temporality and mathematics (Davis, Hersch, *Descartes's Dream*, 1990, pp. 189-201).
[10] This expression, used by (Gurwitsch 1974, p. 44) did not mean "hypostasis" in a theological sense. Elements of nature as "mental creation" also appeared in the terminology of Einstein. See, for example, (Einstein 1973, p. 291).
[11] See a careful explanation of the meaning of this term in (Nesteruk 2003, pp. 112-17; 2004).

[12] It was his last book, *The Crisis of European Sciences* (Husserl 1970), where Husserl undertook a critique of the mathematisation of nature whose inception was associated with the name of Galileo. The topic was later discussed and developed in numerous papers. See, for example, (Gurwitsch 1967), (Kvasz 2002).

## Participatory universe and human agency

As we mentioned above the main ambition of Wheeler's concept is to approach the issue of genesis of "meaning" and hence "reality" of the universe in strictly physical terms. He attempts to explicate this genesis by employing as its primary elements observations and measurements as certain intentional acts of consciousness with respect to the world. The physical happenings which are assumed in a naively realistic view as taking place contingently without being observed and measured, are contraposed to those *events* which were brought irreversibly into being by conscious intentions expressed through observations, so that the traces of presence of these intentions in being cannot be erased (Wheeler 1987, p. 311). These events can be called existential events because they involve human presence and it is in these events that the meaning of what is observed is established. The most difficult philosophical issue is exactly where and how this meaning comes from. Here Wheeler needs to give more precise definition of the human agency as that centre of disclosure and manifestation which provides this meaning.

The human agency is portrayed by Wheeler as a network of observers, who by means of communication establish the meaning of what is called physical reality. Wheeler's thought follows a kind of a reductionist emergent philosophy, by asserting that consciousness is a product of blind physical forces and myriads of particles in the universe. However, the reference to blind physical forces and chance prior to the established human articulation is made as a matter of rhetoric because the universe as the "world of existences" did not exist "prior" to human subjectivity: "observers are necessary to bring the universe into being."[13] The universe thus is a participatory universe; its existence is relational upon the existence of intelligent observers whereas the existence of observers is being relational upon the ingredients of the universe. There is a certain reciprocity between the universe and observers: one cannot exist without the other (DeLaguna 1966, p. 82).

As to the origin of reflecting and articulating consciousness in the universe, Wheeler sincerely believes that science will be able to provide an explanation of the origins of human intelligence in the future (Wheeler 1994[1], p. 307). This corresponds to his implicit desire to treat both intelligence as well as the intelligible image of the universe as emergent properties. The human phenomenon , then, would be an inevitable result grounded in purely natural factors, and the "tangible reality of the universe" would be just natural as well, although of a different, animated or self-reflected order. In one of his famous diagrams, illustrating the transition from the view of the dead mechanical universe to the universe as the world of existences, Wheeler represents the universe as a self-excited circuit, that is as developing through a cycle (closed loop) which excludes reference to any preexistent foundation outside this circuit (Wheeler 1994, p. 293) (see a more sophisticated diagram in, for example, (Wheeler 1988, p. 5)). In both diagrams the self-awareness of the universe through human intelligence, represented by

---

[13] This is a short formulation of the Participatory Anthropic Principle. See (Barrow, Tipler, 1986, p. 22).

Wheeler as the network of observers-participants, completes the "evolution" of the universe in Wheeler's sense as the movement along the closed circuit. In fact, this so called "evolution" cannot be seriously treated as related to the objective pole in the universe, that is as physical or biological evolution as if it were devoid of the human insight. The "evolution" is itself a construction in the course of observer-participancy, whose completion is represented by the intelligent agency explicated by means of the same intelligence which is supposed to be a part of the diagram. Correspondingly there is no way out from this circuit, that is there is no foundation of the circuit outside itself. From a philosophical point of view this means that since the circuit is closed, and the universe receives its explanation from within it, no question on the *purpose* of the universe and its end can be posed in the sense of the material of the *nexus finalis* (that is, as if they existed in objects independently of the human intelligence): purpose and end are just the emergent attributes of the world of existences which pertain to the observer-participancy as human activity. The world's existence and its history according to humanity is explained from within its particular formation and, in reflection, results in a monistic view which does not require any appeal to trans-worldly factors. Wheeler argues that his model of a closed circuit escapes the danger of an infinite regress of causations towards the ultimate substance similar to that of the ancient Greek philosophy (Wheeler 1987, p. 313; 1994[1], p. 300).[14]

However it is not difficult to realise that the notion of underlying physical substance to which one can make an ultimate reference is replaced by the network (community) of human observers who "create" the physical world as constituted reality.

In similarity with existential philosophy and theology Wheeler gives priority to human persons who produce meaning, rather than impersonal substance which is an abstract and impersonal notion. This entails a tacit anthropological assumption about embodiment, which is present in Wheeler's theory of the universe, namely the constitution of intelligent hypostatic observers as unities of sensible bodies and soul which produce the coherent view of the universe: there is no explanation as to why this particular composite was brought into being. The logic of explanation is different, that is, the universe has a dual structure: as an undifferentiated stuff with no meaning (before observers developed) on the one hand, and as sensible agencies and objects with meaning on the other (or, in different words, as the sensible and intelligible) after the network of observers developed the intersubjective meaning of what was observed. This is the reason why the "observers are necessary to bring the universe into being", that is, to transform something initially undifferentiated and non-articulated to things which are sensible and intelligible.[15] This implies, according to Wheeler's logic that the intelligibility of the universe is rooted in the ability of human beings to establish its structures and patterns through communication, starting from some elementary observations-measurements which are described by quantum theory. It is clear that the deposit of intelligibility has been tacitly present in Wheeler's scheme of being from the very beginning. Human observers who explain the universe and their own place in it are already there even if the very scheme of being does not introduce them at a given stage yet. This also makes it clear that in spite of man's "central" position in Wheeler's meaning circuit it is not central enough, for the very existence of this circuit is possible because mankind can transcend its particular place in it and integrate the whole circuit in a single consciousness. This implies that while being a part of the meaning circuit human beings transcend it in the sense that they have an a-priori ability to contemplate the universe as a whole and position themselves in it before they consciously account for their position through the abovementioned diagram. In other words, there is an inherent consubstantiality between human observers and the universe which is not articulated at the initial parts of Wheeler's scheme, but which is tacitly present through the very possibility of depicting those parts of physical reality which have not yet produced humanity. But this consubstantiality has, so to speak, a transcendental character; for it is detected by human consciousness through the next order of reflection. In this humanity exhibits itself as metaphysically infinite creatures, living in the conditions of a physical finitude, the finitude which is constituted by humanity itself from the perspective of the infinite.

[14] Wheeler argues that his approach to understanding the place and role of man in the universe contrasts to the selection mechanism of the many worlds (MW) version of the Strong AP (which assumes pre-existence not only of the visible universe, but also the multitude of other universes) in a sense, that the Participatory AP is "founded on construction" (Wheeler 1987, p. 310). He articulates this contrast as an opposition in views on the place of man in the universe as mediocre versus central: "Life, mind, and meaning have only a peripheral and accidental place in the scheme of things in this view [i.e. MW-Strong AP (A.N.)]. In the other view [that is, Participatory AP (A.N.)] they are central. Only by their agency is it even possible to construct the universe or existence, or what we call reality. Those make-believe universes totally devoid of life are (according to this view) totally devoid of physical sense not merely because they cannot be observed, but because there is no way to make them" (Ibid.).

[15] The place of observer is not to "create out of nothing" in a theological sense, but to act a an ancient god-demiurge who orders the universe from preexistent matter.

Here the analogy with transcendental philosophy can be invoked. Indeed, if Wheeler claims that the observers bring the universe into being, including its space, time, etc. (Wheeler 1988), then one can reasonably ask: where do human observers do this *from*, if there is no preexistent space and time? This question is reminiscent of the famous Kantian affirmation that human being is *phenomenon* and *noumenon* at the same time. On the one hand human beings as biological organisms are in space and time. On the other hand, according to Kant, space and time represent transcendental forms of sensibility as the necessary conditions for human perception of the physical bodies. This means that because human beings constitute space and time from the depths of their transcendental *ego* which is eidetically "prior" to any particular form of physical embodiment, one can conjecture that they inhere in "something" which is beyond and prior to space and time and which, at the same time, contains in itself the potentiality of being explicated in terms of space and time.

Wheeler attempts to claim that the meaning of space and time, as well as all other attributes of the universe, is constructed through observership-participation in acts of cognition resembling quantum measurements. A philosopher, in the style of Kant, would object to this by saying that the sense-data alone can not constitute the notions of space and time and that, vice versa, the ordering of the sense-data can only be done in rubrics of space and time which are a-priori forms of sensibility. Whereas a phenomenological stance would be that space is not pre-existent and objective "out there", originating from subject's passive contemplation of it, but in terms of subject's comportment "in" it. This, so called, *attuned* space becomes an initial instant and a medium of disclosure of that "objective" space through relation to which this subject is constituted as a corporeal existent in space. However, this relationship is manifest of a paradox similar to that of the container and of the contained put in an interrogative form: how can one grasp the relationship of a particular being (subject) as if 'in' space when this being is essentially constituted by being 'over against', and hence beyond space? (Ströker 1965, p. 15). This once again brings us to the Kantian stance on human being, as being simultaneously phenomenon and noumenon: on the one hand space is an a-priori form of sensibility which allows a subject to order its experience; on the other hand this form of sensibility is unfolded not from within that space which is depicted by it, that is it comes from beyond any possible spatial presentation of experience.

What is obvious, however, is that the constitution of space, first of all of the attuned space, is intertwined and not detachable from the fundamental aspect of human embodiment or corporeity. Embodiment or corporeity manifests itself not as a system of some biological processes nor as simply a body animated by the soul, nor even a simple unity of both of them. It is not also a lived body (*corps-sujet* in a sense of G. Marcel); it is a living being in relation to other beings and to the world, in whom this relation is announced and articulated by the way of its sense-reaction and its comportment, or its action in situation. In this sense the constitution of space in all its varieties (from attuned space of immediate indwelling to mathematical space of the universe) represents the modes of explication of embodiment or corporeity of humanity through which it interacts with the world. Thus the lived body entails a kind of lived space which bears the character of self-givenness "in the flesh". In other words, the initial point of any discourse on corporeity and associated spatiality implies knowledge as presence "in person" or "in the flesh" as a mode of givenness of an object in its standing in front of the functioning corporeity.

The question of embodiment becomes acutely important in Wheeler's scheme of origin of the universe as the world of existences. Indeed, what is important is that the network of observers is the community of embodied creatures, and this embodiment *per se* reflects the pre-existing physical conditions which are not subject to considerable change during the span of the human civilisation. In other words, one can assume together with Wheeler that these conditions of embodiment as statements of physics have not been always articulated, but they have been implied, so that any physics which follows from a cognitive acquisition of the world is prone to contain the conditions of embodiment as transcendental conditions for the explicability of nature. It is in this sense that the famous Weak Anthropic Principle (Weak AP) can be taken into account: what we observe may be restricted by the conditions necessary for our embodied existence as observers. Then the unfolding process of Wheeler's articulation of the world contains in itself the very possibility of existence of those creatures who articulate this world. In spite of a contingent path of articulation of nature, which is related to the history of the sciences, this contingency contains an element of *necessity*: this articulation and constitution takes place only in relation to that bulk of being which is unconcelad through the conditions of embodiment. The famous thesis of Protagoras that man is the measure of all things, seen through the eyes of Heidegger (Heidegger 1987, p.91-95), gives strength to our assertion that the constitution of reality according to Wheeler, being an open-ended process, is still human-centred because the very explicability of the world through the chain of questions and answers is subjected to the condition of its origination in embodiment. It is important to realise that neither the Weak AP, nor the participatory genesis of Wheeler make an explicit link of the limits of embodiment expressed in physical terms to the limits of cognitive methods of exploration of reality (as was famously proposed by Kant through reference to Euclidian Geometry and Newtonian physics). In terms of exploration of physical reality one can either gain access to processes beyond the scales of embodiment through technology, or even transport the very conditions of embodiment on spaceships to some hostile terrain in order to gain knowledge of that which is beyond the Earthly horizon of embodiment. In this sense the Weak AP or the Participatory strategy of Wheeler state nothing of the restrictions on the methodology of research, or in different words, they avoid any commitments to the limits of possible knowledge in the sense of its methods. In this sense they both are more flexible in comparison with the Kantian dogmatic position. However the Kantianism remains in both Weak AP and Participatory AP in a hidden and more subtle way through the implicit teleology which pertains to the very way nature becomes unconcealed and explicable in the conditions of embodiment.

In order to make the ambivalent position of the human observer (noumenon/phenomenon) more explicit, one might place our discussion back in the Kantian frame of mind. Space and time as being constituted by intelligible observers in Wheeler's scheme of things can be said as being brought into being by transcendental observers whose existential centre relates to the physical world but is not exhausted by it. In this sense the genesis of reality, or its constitution as articulated in consciousness, appeals to the realm of the intelligible. Space and time as articulated notions, as well as the whole universe appear as mental images of reality, as ideas which have a precarious relation to physical reality. By learning the lessons from Kant, it can then be anticipated that any attempt to provide a coherent picture of the 'genesis' of the concept of the universe, that is to speculate about its ultimate grounds on the physical level, will inevitably lead reason to an antinomian difficulty. This consequently places the notion of the community of observers which gives meaning to the universe in an ambivalent position of being *in* the world, and at the same time not *of* the world.

Indeed, one can conjecture that the thesis of Wheeler's Participatory AP, namely that "observers are necessary in order to bring the universe into being", makes the notion of the network of observers in Wheeler's concept similar to the idea of an absolutely necessary being that appears in the fourth Kantian antinomy so that Wheeler's proposition can be reformulated as a *participatory anthropic antinomy* (See also (Nesteruk 1999, p. 83; 2003, p. 225)):

*Thesis*: The network of intelligent observers understood in a transcendental sense as existing in whatever relation to time and space is absolutely necessary for the visible universe in space and time to be brought into being.

*Antithesis*: The existence of the visible universe with spatiotemporal attributes is not contingent upon the existence of the network of intelligent observers (understood in a transcendental sense) as its cause either in the visible universe or out of it.

This antinomy indicates the dichotomy in the ontological status of the network of intelligent observers as having a specific location of their embodiment related to the physical conditions of survival and, at the same time, as transcending these specific places and establishing the sense of space and time out of some originary propensities that define the observers as intelligible and consciously non-local beings. This seeming paradox, which represents a particular reincarnation of the paradox of human subjectivity mentioned above, contributes to the constitution of human observers as composites of the corporeal and intelligible whose contingent facticity cannot be accounted for, remaining thus a primary metaphysical mystery. This cascades up to the mystery of the universe as the constituted *World of Existences*: since the origin of human personhood can hardly to be reduced to the results of impersonal chances and necessities in physical factors and causality, its presence in the universe can be treated as an *event* which can be called the *humankind-event* (Nesteruk 2003, pp. 194-200). This event is indeed formative for the universe to exist, that is to be manifest and disclosed in human personhood. Then the process of constitution of the universe in Wheeler's participatory scheme reveals itself as the *enhypostasization* of the universe within the humankind-event, that is the universe itself becomes no more than an *event* related to the history of humanity, a flash of the universe's self-consciousness depicted in Wheeler's writings by a diagram of the human eye emerging in the bold letter **U** (symbolising the universe) which itself is the formation of this eye (Wheeler 1994, p. 293).

Some other philosophers formulated a similar "eventiality" of the universe by referring to a communal character of events of knowing. In P. Heelan's terms, "the phenomenon takes 'flesh' in the world differently because its 'flesh' is determined only as a consequence of decisions taken by local and historical communities of expert witnesses"(Heelan 1992, p. 58).[16] It is in this sense that the articulation of the past of the universe is an event within the life-world of a particular community loaded with a sense of the community's lived past and of decisions to be made in the future. As Heelan points out, "it is not the case that every historical event is also an event of a scientific kind..., but when the local community is one of expert witnesses, then the scientific data produced by that community are also historical events in relation to that community" (Heelan 1992, p. 66). H. Margenau argued in a similar way that "physical reality" is best defined as the totality of all valid constructs. In this approach the universe is defined not as a static, but as a dynamic formation: "...the universe grows as valid constructs are being discovered. Physical entities do not exist in a stagnant and immutable sense but are constantly coming into being" (Margenau 1944, p. 278).[17]

[16] The metaphor of 'flesh' is borrowed from M. Merleau-Ponty.

Wheeler, as well as the others, attempts to make a point that the sense of physical reality is not a pregiven compendium of laws and facts and that it originates in the constitution of this reality through formation of meaning of the universe through communication in the network of observers. Wheeler does not give any specific model of how to deduce the meaning from the forms of intersubjectivity, but his "reconstruction" of logical steps involved in making physical theories articulates the fact that in spite of the idealisation and mathematization with which modern physics operates, there is still a level of understanding which itself cannot be described through them. In other words, physics cannot account for its own facticity only through physics itself: one needs to appeal to such an order of reality by reference to which physics at least receives its interpretation. This implicitly points out to the mundane fact that the scientific advance, despite its complex language, is ultimately rooted in the primary experience of the world, in that which phenomenology calls the *life-world*.[18] This implies that the sphere of human subjectivity as immediately given and irreducible to any scientific analysis is assumed by Wheeler to be present in order to develop from within it the articulated picture of the world with a special language and mathematics. Wheeler's conjecture that the whole edifice of modern physics can be reduced to a simple quantum principle, "it is from bit" (Wheeler 1994, p. 295), shows that there is still a certain level of reality behind these "bits", which constitutes the meaning of any sequence of those bits and this reality is the mystery of embodied human consciousness which endows reality with meaning. However, the observers who possess the ability to articulate the external world, are incapable of comprehending the very possibility of acts of consciousness which are responsible for the articulation of the world. Can physics explain them in naturalistic terms? In spite of a heroic attempt of Wheeler to propose a scheme for elucidating this problem, his intellectual construction of the participatory universe demonstrated with a new force that the main philosophical mystery of human intentional consciousness and its engagement with the world still remains (Cf. Marcel 1965, p. 24).

In spite of a philosophical scepticism with respect to Wheeler's attempts to give a physicalistic model for the genesis of meaning of the universe as a process of mutual interaction between the network of observers and their physical environment one should admit that they contributed in a non-obvious way to the rearticulation of the life-world as that primary existential milieu which lies in the foundation of scientific articulatiuons of reality. The search for the foundations of the universe, as well as the foundations of physics leads inevitably to the recognition of the centrality of existential "immediate here" and "immediate now" from which the whole grandeur of the universe (as the world of existences) comes into existence by "contracting its existing" in life of human observers (Cf. (Levinas 1978, pp. 82-85; 1987, pp. 42-44). This confirms a general philosophical conviction that science contributes to the understanding of life and humanity, for "the whole universe of science is built upon the world as directly experienced, and if we want to subject science itself to rigorous scrutiny and arrive at a precise assessment of its meaning and scope, we must begin by reawakening the *basic experience of the world of which science is the second-order expression*" (Merleau-Ponty 1962, p. ix (emphasis added)). That which was formulated by M. Merleau-Ponty in the context of existential philosophy has been renewed in Wheeler's metaphysical extension of physics: namely to remind physicists that all their notions are ultimately inherent in the very specific place human beings occupy in the world which they attempt to articulate. However, the nature of human subjectivity and intersubjectivity, for obvious philosophical reasons [19], did not enter as a constitutive principle of physical explanation. The mystery of human intelligence was recognised as pivotal for developing an articulated picture of the world, but, at the same time consciousness and intelligence were treated in a reductionist sense as products of physical and biological evolution. If Wheeler's model of the participatory universe is assessed through the eyes of an existential stance on the primacy of the sphere of human subjectivity, expressed, for example, in the words of Merleau-Ponty, quoted above, then obviously the universe as a self-exited circuit must require for its ontological assessment one crucial element: the presence of conscious insight overlooking this universe from above and beyond.

---

[17] Margenau anticipated that many scientists would disagree with such an attitude because they maintain a faith in the convergence of the system of the entire set of physical explanations which would deliver them an ideal of their aspirations, that is a unique and ultimate set of constructs for which would reserve the name 'reality'. However he points out that this belief in convergence in question is problematic because it is not capable of scientific proof. ((Ibid). See also (Margenau 1977, p. 76) The situation in modern cosmology, where the ever increasing set of theoretical constructs reveals the components of the matter content of the universe which escape physical description points exactly to the danger of idealisation of the scientific description of the universe: the more details we know the less we understand the entirety.

[18] The concept of the life world was introduced in Husserl's Crisis (Husserl 1974) and was a matter of vast discussion by phenomenological philosophers. See for a recent review (Steinbock 1995).

In order to make the latter thought clear, one can suggest a graphical interpretation of three typical cosmological diagrams which pretend to catch the unity of physical reality at different spatial scales and other physical parameters. Indeed at the figure in the Appendix we have three different representations of such a unity: in the upper left corner there is an image of the so called *Uroboros*, symbolising the interconnectedness of physical entities at different spatial scales of the universe (Primack 2006, p. 160). Humanity's consubstantiality with the universe is depicted at the bottom centre of this diagram. In the second diagram, below, there is a display of various objects in the universe according to their sizes and masses (Barrow 1999, p. 32). Once again humanity is positioned as a mediocre physical formation at the centre of this diagram. Both these diagrams are presented as static formations which do not reflect any processes or genesis of these diagrams, that is their epistemological constitution. Correspondingly, any attempt to predicate on their basis that humanity is physically insignificant in the universe will be philosophically weak because both diagrams are mental creations and humanity is present in them not only through its insignificant position but above and beyond all its elements as an articulating consciousness. This is depicted by positioning the human subject outside the diagrams while retaining the traces of its physical embodiment. If now one compares the *Uroboros* and the size/mass diagram with Wheeler's graphical representations of the participatory universe (the right-hand side of the same figure) presented through a genesis of physical properties of the world towards its intelligibility (Wheeler 1988, p. 5), then the difference is clearly seen: the diagram attempts to encapsulate a temporal aspect in formation of the overall picture of the universe and make manifest that the universe is an accomplishment because the human phenomenon in it is itself an accomplishment. However the presence of human beings as forms of biological life does not entail with any physical necessity the presence of intelligence. Correspondingly the diagram of the closed-circuit universe is not just an accomplishment of physics, it is a mental accomplishment which is contingent upon intelligence, which is, through embodiment, a part of the diagram and, at the same time, something outside of it. In other words Wheeler's diagram presupposes for its own existence the presence of intelligence which creates this diagram: humanity appears in it as the centre of disclosure and manifestation. The physical genesis depicted in this diagram requires a reflecting consciousness which has propensities which do not simply follow from the chain of physical causations. In this sense the genesis of physical properties leading to the fulfilment of the necessary conditions for observer's existence does not entail the fulfilment of the sufficient conditions which justify intelligence and the way this intelligence approaches reality through the logic of questions and answers suggested by Wheeler. Once again, what is missing in Wheeler's diagram is the presence of the subject for whom this diagram makes sense. And this subject is above and beyond this diagram (depicted at the centre), in that directly experienced world, given to humanity in its embodied consubstantial constitution so that the diagram itself, phrasing this in the language of Merleau-Ponty, is *the second-order expression* of this world.

[19] The fundamental problematic character of any philosophical enquiry into the nature of human consciousness is expressed in modern terms through a concept of the so called "negative certitude" meaning that the facticity of consciousness can only be approached with certainty in negative terms, that it is certain that its mystery can only be predicated in terms of that which is not this consciousness (see (Marion 2010, pp. 21-86)).

From a philosophical point of view there is a gap in Wheeler's reasoning on the universe as an emergent meaning circuit, for there is no explanation as to why the intelligent observers, who reveal the intelligibility of the entire universe, are possible at all. In other words, why the universe entails the transcendental conditions of its own explicability. Here Wheeler invokes a kind of a theological reference by affirming that the whole situation in modern science completely changed the problem of creation as the problem of the relationship between man and God. If in the traditional theology human observers were treated as created by God in his image and thus having an innate ability to articulate the whole creation, in contemporary science these observers are treated as a result of the natural development of the world, so that the articulation of the world is part of the ongoing process of development of humankind. To accentuate this contrast Wheeler in one of his texts refers to an old legend of the dialogue between Abraham and God (which manifested the relationship between man and God) and says that "in our time the participants of the dialogue changed. They are the universe and man" (Wheeler 1994, p. 128). In the same passage he imitates the dialogue between the universe and man as an act of personifying the universe through the sequence of questions and answers. The universe acquires a sort of "intelligibility" as its "natural artefact".

In conclusion, the main interesting result of Wheeler's attempts to sketch the "physics of meaning" was the rediscovery of the issues of the life-world. Physics has sense as long as it has meaning, which was assigned to it by human beings. This means that physics is essentially human, as well as the universe constructed through physics, and represents an intentional correlate of human intersubjectivity, so that it is given to us in so far as it contains us. The world of classical physics which was deprived of its inward existence in subjectivity, receives back its existential meaning through the metaphysical extension of some propositions of quantum physics. In spite of the fact that there is no direct reference to the Divine as the ultimate source of human intelligibility in the world, there is a reference to the otherness of the world which is implicitly made in Wheeler's models through posing a fundamental question of meaning: why is the universe?

## Implicit Purposiveness in Wheeler's Participatory Universe

Finally we want to discuss a teleological aspect of Wheeler's thought. For the universe to become the World of Existence this same universe must have conditions for the emergence of intelligent life and thus its explicability. Correspondingly if the World of Existences, as it seems from Wheeler's writings, is that high term in the overall chain of transformations in the universe, so to speak its "ontic" goal, then the presence of the human intelligence in the universe is somehow implanted into this goal. When the Participatory AP asserts that *observers are necessary* for the universe to come into being it effectively states that the development of the universe must have a *necessary condition* for the emergence of human beings. Here one can ask a question as to whether there is an implied teleology in Wheeler's view of the universe, where teleology is referred to a definite material pole, that is to that physical state of the universe when human life is possible. Probably one must give a negative answer to this question because the necessary conditions for emergence of biological life in the universe do not automatically guarantee the emergence of intelligence. The sufficient conditions for the emergence of intelligence in the universe are not subject to the physical description. Correspondingly the seeming teleology of Wheeler's account for the genesis of the World of Existences is not related to the material goal of the universe's development; it rather relates to another teleology, associated with the explicability of the universe by human agency. In fact, in Wheeler's case, this explicability is closely connected with the constitution of meaning of the universe: to constitute the universe as the World of Existences one must establish the meaning of things in this universe. Thus the whole pattern of Wheeler's reasoning when he invokes an analogy with the quantum questions and answers points not to the material pole, or a result of its constitution (for it seems to be an open-ended process) but rather to the strategy or methodology of the scientific quest for meaning of the universe, a particular way of interrogation of nature and its outcome delimited by this way (See (Nesteruk 2012[31]). The objective of physics is to explain the universe; the *telos* of this explanation is not something which pre-exists this explanation, but that remote epistemological ideal, a supposed mental accomplishment, which would correspond to an imagined convergence of different strategies of explanation and correspondence rules. In spite of a principal impossibility of stating even roughly the possible ideal pole of such an explanation, there is one teleological example which makes it possible to elucidate the sense of what Wheeler implied in his idea of the Participatory universe. This example relates to the Big Bang, the ultimate beginning and origin of all things in the universe, including human intelligence itself.

The notion of the Big Bang was at the centre of Wheeler's discussions on the nature of space and time as that epistemological boundary beyond which physics cannot proceed. He also drew the conclusion that since the notions of space and time loose their physical meaning in the singularity of the Big Bang, they must be considered mutable ingredients of physics subject to constitution. Correspondingly the Big Bang itself is not an immutable material pole associated with the origin and beginning of all things, but a construction whose anomalous properties point not so much to the limiting capacities of physicists to deal with the questions of origin, but rather to a specific way in the acquisition of reality (through the logic of questions and answers) which leads to the constitution of what is meant by the Big Bang. It is remarkable, however, that the process of constitution of the universe, as being directed in the future of the historical time associated with observers, encompasses all temporal aeons of the universe, including its allegedly existing past. This means that not only our actual present is subject to constitution, but what is aimed to be the past and future of the universe is constituted by the human observers and thus their ontological status becomes ambiguous. The Big Bang, for example, appears to be also a mental construct dealing with the alleged past of the universe, but only through references to here and now, because its theory is constructed upon observations made here and now and progressing to the future. Correspondingly, for Wheeler, the question of the physical existence of the Big Bang has no sense if it is not placed in the context of how it is constituted and articulated by human observers here and now. He expresses this conviction by posing a question: "Is the term

'big bang' merely a shorthand way to describe the cumulative consequence of billions upon billions of elementary acts of observer-participancy reaching back into the past [..] ?" (Wheeler 1994[1], p. 128; Cf. Wheeler 1985, p. 387). Elsewhere Wheeler generalises this thought applying it to the constructed temporality of the past: "The 'past' is theory. The past has no existence except as it is recorded in the present. By detecting what questions our quantum registering equipment shall put in the present, we have an undeniable choice in what we have the right to say about the past" (Wheeler 1985, p. 366; 1988, p. 13).[20] The acts of observer-participancy intend to reach to the past of the universe, whereas their conscious dynamics is constantly directed to the future. Certainly, according to Wheeler, there is no sense to enquire into the "objective" sense of the universe before or beyond the intelligence emerged; as expressed in a philosophical context by Christos Yannaras, "even the formation of the universe "before" the appearance of its human cognition does not destroy the character of being invited-to-relationship of the universe's referentiality. For the "before" and "after" are by-products of the relationship between humanity and the world, the only relationship that constitutes an existential fact and whatever "pre"-required evolutionary process is needed for its realisation" (Yannaras 2004, p. 138) (Cf. (Wheeler 1975, p. 17)).

If the Big Bang is constituted in the ongoing process of exploration of the universe, the whole issue of the initial conditions of the universe as if they were once and forever set from the "outside" of the universe looses its objectivistic sense, because the whole history of the universe is constructed by humanity from its present state so that the past of the universe is seen only in the perspective of the ever moving present and the ultimate point in the past, the origin of all, can then be grasped as a limiting point of humanity's knowledge not only as a boundary of its present state of understanding but as an ideal aspired for through the movement of knowledge to the future, that is as it *telos*. In this case the notion of the Big Bang functions in human knowledge as a limiting point for any historically given state of knowledge, but, at the same time this limit as being extended through the progress of science becomes the ground of its motivation and aspiration explicating not only the Big Bang as a remote physical pole, but also explicating the evolving epistemology of the enquiry in the foundation of the facticity of all, including the very enquiring consciousness. Here we inevitably come to an interesting and counter-intuitive conclusion that the Big Bang, as an allegedly physical pole in the origin of the visible universe, turns out to be the *telos* of scientific explanation, as its ultimate goal to see the origin of the varied display in the universe in the unity of "all in all".[21]

In practically all papers related to the genesis of the World of Existences Wheeler promotes an idea of the cosmological singularity or the Big Bang which has demonstrated to us that all classical laws of physics as well as its basic notions disappear at the singularity, so that the "ultimate, underlying" reality cannot be described in terms of physical laws and categories at all: there is a law, that there is no law. Elsewhere the situation with the impossibility to ascribing to the Big Bang spatio-temporal attributes and at the same time the fact that it is the Big-Bang which supposed to give "beginning" to space and time in their facticity, was qualified as that the Big Bang notifies in theory the "existential" fact of the uniqueness and concreteness of the universe without entailing the materiality of its existence (Yannaras 2004, p. 107). Indeed, if the Big Bang, according to Wheeler is the construction, that is constituted through the relationship between the world and humanity, then what is the objectivistic and material status of this construction: does the phenomenality of the Big Bang falls into the category of the physical "out there"? Yannaras poses a question in an even more radical form: if the Big Bang is the metaphysical concept which presupposes an obligatory exit from succession of before and after, and also from every dimensional location, does it entail the exist from the presupposition of the existent? (Yannaras 2004, p. 105) Wheeler's answer is that it does not: the Big Bang can be attributed the status of the existent but in a strictly constituted sense in the same way that he advocates that the laws of classical physics are constituted by us.

[20] This thought must be placed into an even more general conviction that in the ultimate scheme of things there is no time or temporality at all. Temporality is a human construction: "The word Time came, not from heaven, but from the mouth of man, an early thinker, his name long lost. If problems attend the term, they are of our own making" (Wheeler 1994[2], p. 6).

[21] This point was developed in (Nesteruk 2008, 2012[1]).

It is not difficult to conjecture that the only "real" law which drives physics is the "law" that the universe must be explicable. It is impossible to deny this requirement for explicability even in Wheeler's thought, for, in fact, all his edifice of dealing with the genesis of physics is to advocate the explicability of the universe whatever philosophical orientation taken. In this sense it is this explicability which becomes the ultimate *telos* of the whole complex of human observers – the universe. The maxim of teleology, if one uses Kant's terminology, ordains in Wheeler's scheme of things the use of some established physical law-like strategies for giving more precise details of the genesis of physical objects.[22] There is an implicit purposiveness in the closed circuit established between observers and physical reality which ultimately proceeds from the nature of observers as human intelligent beings endowed with the purposiveness of any actions. This purposiveness, in order to avoid any classical and unfashionable teleology related to the physical development of the universe, must presuppose an extra-physical character. Seen theologically, the purposiveness can proceed from the Divine image, set up exactly for the purpose of bringing the universe to union with God through an integrate knowledge of it (Nesteruk 2003, p. 230). If this theological stance seems to be unsatisfactory and one becomes inclined towards a materialistic reductionism, attributing the purposiveness of explanation as being implanted in physical reality, then the alleged purposiveness of the universe brings us to the question of its subject: who is that intentional agent for whom the universe has a purpose? It is not difficult to see that the idea of the Divine subjectivity enters the scheme of things at a different level: the Participatory AP in this case becomes similar to that version of the Strong AP which postulates that the universe must have human agencies as its product at a certain stage of its development. If this is true, then the difference between the Participatory AP and the Strong AP, which was so emphatically advocated by Wheeler become blurred. Our analysis thus unfolds the most important and metaphysical point to be made on Wheeler's ideas, namely the mystery and precarious status of human agents, observers-participants in what concerns the origin of their purposive actions which are in the foundation of knowledge of the universe. It seems that Wheeler's hope that human intelligence and correspondingly purposiveness as such will eventually become a subject of explanation by physics remains in vain, for the basic question of the facticity of the human intelligent agency, in spite of all reductionist hopes, remains unanswered. Purposiveness is a human aspect of existence and one can hardly believe that physics, being purposive activity, can explain the emergence of this purposiveness out of itself. What is important, however, is that the existence of life and intelligence, being an "experiential fact" and determining the lines of scientific enquiry provides "the unlimited informative value for the universe and its laws" (Yannaras 2004, p. 117).

[22] See more details on this issue in (Nesteruk 2012[3]).

**Appendix: Humanity as the centre of disclosure and manifestation of the universe**

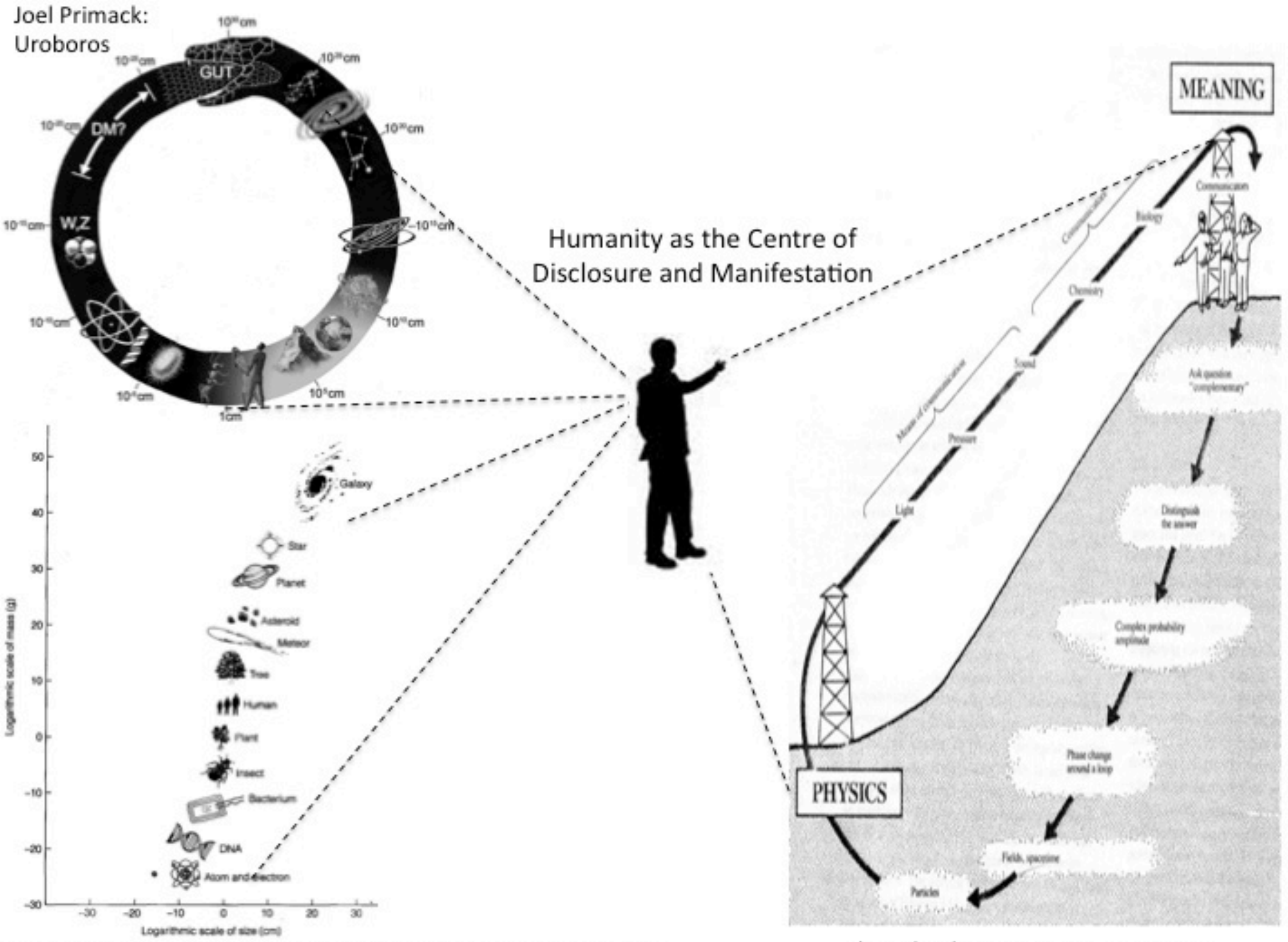

John Barrow: the unity of the universe through masses and scales

John Wheeler: Participatory universe

**Acknowledgements:**
This publication was made possible through the support of a grant from the John Templeton Foundation. The opinions expressed in this publication are those of the author and do not necessarily reflect the views of the John Templeton Foundation. I would like to express my feelings of gratitude to George Horton and Christopher Dewdney for reading the manuscript and making helpful suggestions.